\documentclass[twocolumn]{aastex631}

\usepackage{graphicx}
\usepackage{dcolumn}
\usepackage{bm}
\usepackage{color}

\begin{document}

\title{Ignition of carbon burning from nuclear fission in compact stars}

\author[0000-0001-7271-9098]{C. J. Horowitz}
\affiliation{Center for Exploration of Energy and Matter and
                  Department of Physics, Indiana University,
                  Bloomington, IN 47405, USA\email{horowit@indiana.edu}}




\begin{abstract}
Type\textrm{--}Ia supernovae (SN Ia) are powerful stellar explosions that provide important distance indicators in cosmology.  Recently, we proposed a new SN Ia mechanism that involves a nuclear fission chain reaction in an isolated white dwarf (WD) [PRL {\bf 126}, 1311010].  The first solids that form as a WD starts to freeze are actinide rich and potentially support a fission chain reaction. In this letter we explore thermonuclear ignition from fission heating.  We perform thermal diffusion simulations and find at high densities, above about $7\times 10^8$ g/cm$^3$, that the fission heating can ignite carbon burning.  This could produce a SN Ia or another kind of astrophysical transient.   
\end{abstract}

\keywords{Type Ia supernovae (1728); White dwarf stars (1799)}
\section{Introduction}\label{Sec.Intro}
Type\textrm{--}Ia supernovae (SN Ia) are great stellar explosions that provide important distance indicators in cosmology \cite{Abbott_2019,SN_cosmology,Sullivan2010}.   They can be observed at great distances and appear to have a standardizable luminosity that can be inferred from other observations \cite{1996ApJ...473...88R,Phillips_1999,Goldhaber_2001,Phillips2017,Hayden_2019}.  This allows a precise determination of the expansion rate of the universe known as the Hubble constant.   Nevertheless, there is still some uncertainty as to the  SN Ia explosion mechanism and their progenitor systems.

Traditionally, SN Ia are thought to involve the thermonuclear explosion of a C/O white dwarf (WD) in a binary system. Here the companion is either a conventional star (single-degenerate mechanism) or another WD (double-degenerate) \cite{2012NewAR..56..122W,hillebrandt2013understanding,RUIZLAPUENTE201415}.  
Recently we proposed a new SN Ia mechanism that involves a nuclear fission chain reaction igniting thermonuclear carbon burning in an {\it isolated} WD \cite{PhysRevLett.126.131101,fission2}.  Alternative mechanisms to ignite isolated WD include dark matter interactions \cite{PhysRevLett.115.141301,PhysRevD.105.083507} or pycnonuclear fusion of impurities \cite{10.1093/mnras/stv084}.

Our model involves three stages.  In the first stage, phase separation upon crystallization produces an actinide rich solid that could support a nuclear fission chain reaction.  In a WD, melting points of the chemical elements scale as their atomic number $Z^{5/3}$.  Actinides have the highest $Z$ and may therefore condense first.  The composition of the first solids is discussed in \cite{PhysRevLett.126.131101}.  

The concentration of actinides by chemical separation in a WD is similar to the formation of uranium rich veins on earth.  Not only has uranium been purified by natural processes on earth, natural chain reactions have occurred.  The Oklo natural nuclear reactors operated 2 Gy ago in very rich Uranium deposits in Africa \cite{GAUTHIERLAFAYE19964831,PhysRevLett.93.182302,10.2307/24950391}.  

In the second stage of our model, a chain reaction occurs in a WD.   Nuclear reaction network simulations of this stage were presented in \cite{Fission_network} where it was found that the reaction proceeds very rapidly.  Fertile isotopes such as $^{238}$U or $^{232}$Th can burn via a two step process where a neutron is captured to produce an odd A isotope that fissions after absorbing a second neutron.  As a result, a large fraction of the initial U and Th fissions producing significant heating. 

In the third stage, fission heating ignites carbon burning and initiates a SN Ia or other astrophysical transient.  In this letter we present the first simulations of this stage.   We find that the fission heating can initiate carbon burning if the density is high enough.  For context, our mechanism is similar to a hydrogen bomb.  The Classical Super is an H-bomb design that uses heat from an atomic bomb to ignite hydrogen isotopes \cite{Ford}.  The Classical Super likely fails because too much energy is lost to radiation.  In contrast modern weapons may use radiation to first compress the system to higher densities where there is less energy loss.  Thermonuclear ignition may be easier at high densities.  Therefore, we  explore ignition for different WD densities.

\cite{1992ApJ...396..649T} discuss ignition in terms of heating at least a trigger mass $M_{trig}$ of material.  $M_{trig}$ is estimated from the mass in a sphere of radius equal to the carbon burning flame width $\delta$. $\delta$ decreases with density roughly as $\rho^{-5/3}$ so that $M_{trig}$ decreases rapidly as $\approx\rho^{-4}$.  
In our model, the mass of an actinide rich crystal likely exceeds $M_{trig}$ at high densities.

In Sec. \ref{Sec.form} we review the results of \cite{PhysRevLett.126.131101} for the size of the initial actinide rich crystal.  Next, we extend the fission reaction network simulations of \cite{Fission_network} to higher densities.  We then describe our thermal diffusion simulations.  Results for carbon and oxygen ignition are presented in Sec. \ref{Sec.Results}.  We end by discussing possible implications and conclude in Sec. \ref{Sec.Conclusions}.     

\section{Formalism}\label{Sec.form}

{\it Actinide rich crystallization:} As a WD cools it eventually crystallizes.  However just before the main C and O components start to freeze, higher $Z$ impurities may condense since they have much higher melting temperatures.  This process is described in \cite{PhysRevLett.126.131101} where the crystal is assumed to grow by diffusion until a chain reaction is started by a neutron from spontaneous fission.   The crystal mass $M_{pit}$ forms the fission core of our simulation and we refer to it as the pit in analogy with nuclear weapons. $M_{pit}$ is estimated by setting the time to grow by diffusion equal to the time between neutrons.   Extending the analysis of  \cite{PhysRevLett.126.131101} to other densities gives the results in Table \ref{Table2}.  The pit mass is seen to increase slowly with density  $M_{pit}\propto \rho^{3/10}$.

\begin{table}[tbh]
\caption{\label{Table2} Actinide rich crystal (pit) mass $M_{pit}$ and radius $r_{pit}$ for different densities $\rho$.}
\begin{tabular*}{0.29\textwidth}{c c  c } \hline
$\rho$ (g/cm$^3$) & $M_{pit}$ (mg) & $r_{pit}$ (cm)  \\ \hline
$10^8$& 10 & $3\times 10^{-4}$  \\  
$8\times 10^8$ & 20 & $2\times 10^{-4}$\\
$3\times 10^{9}$ & 30 & $1.3\times 10^{-4}$\\
\hline \hline
\end{tabular*}
\end{table}

\begin{figure}[tb]
\centering  
\includegraphics[width=0.5\textwidth]{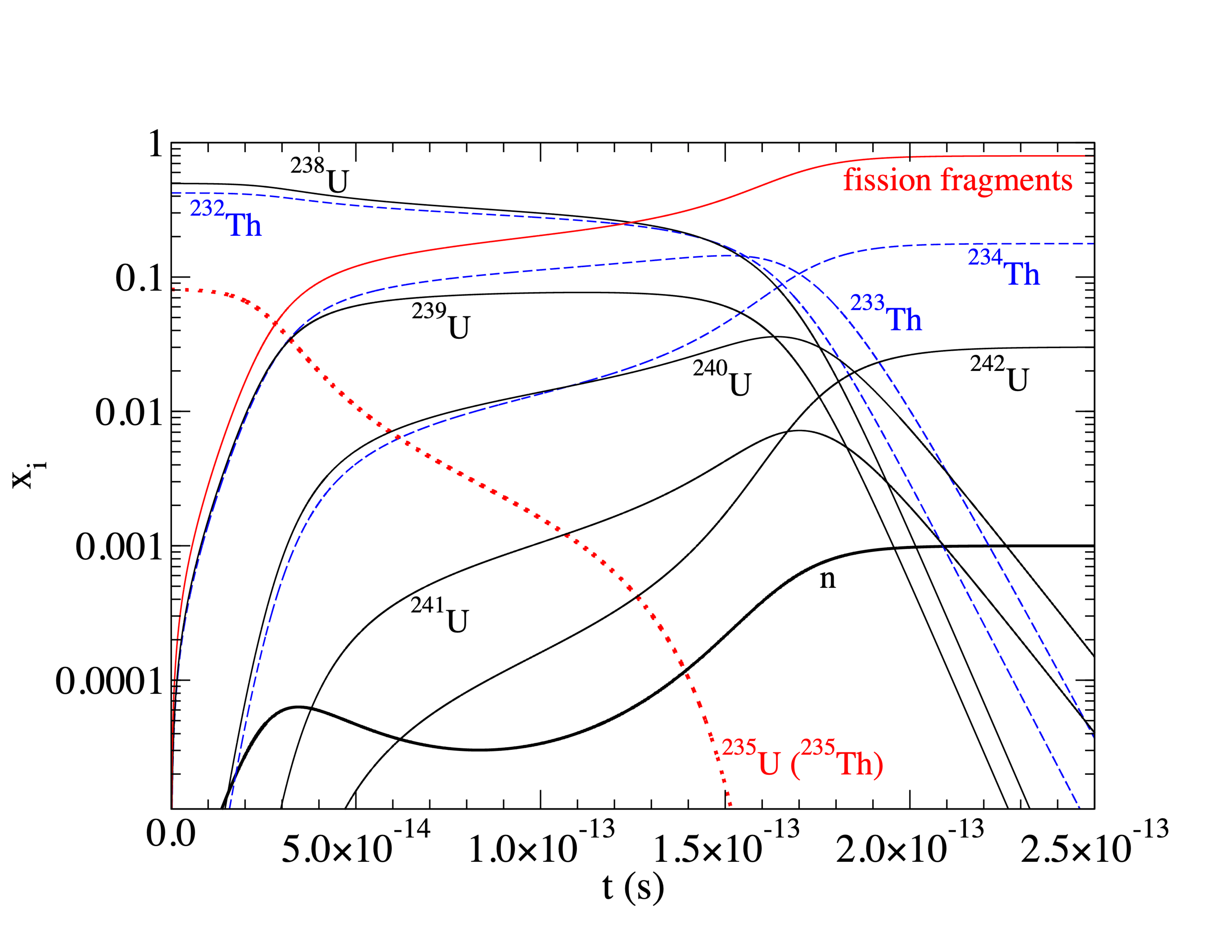}
\caption{\label{Fig1} Abundance of U and Th isotopes (by mass) versus time during fission chain reaction at a density $\rho=8\times 10^8$ g/cm$^3$.  Adapted from Case B of  \cite{Fission_network}. }	
\end{figure}
 
\begin{figure}[tbh]
\centering  
\includegraphics[width=0.5\textwidth]{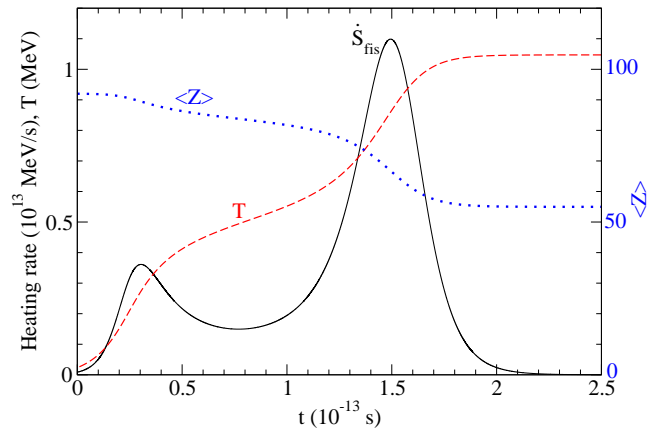}
\caption{\label{Fig2} Fission heating rate per baryon $\dot S_{fis}$ versus time (black solid curve) at $\rho=8\times 10^8$ g/cm$^3$.  Red dashed curve shows temperature at $r=0$ of the thermal diffusion simulation shown in Fig. \ref{Fig3} (c).  Blue dotted curve shows average charge $\langle Z\rangle$ of ions in the pit, to be read from the right hand scale. \explain{Improved labels on curves.}}	
\end{figure}

{\it Fission chain reaction:} This crystal, if critical, will undergo a fission chain reaction. Nuclear reaction network simulations were presented in \cite{Fission_network}, see Fig. \ref{Fig1},  where the fission heating rate per baryon $\dot S_{fis}$ was calculated, see Fig. \ref{Fig2}.  The initial composition included some Pb in addition to U and Th.  Pb is essentially inert during the reaction but increases the heat capacity and therefore acts to dilute the fission heating and reduce the maximum temperature.  However, the composition of the initial solid is uncertain.   To explore ignition most simply, we now consider a composition identical to \cite{PhysRevLett.126.131101} but without Pb
.  The composition shown in Fig. 1 and the fission heating in Fig. 2 is other wise identical to Case B of \cite{Fission_network}.  If Pb is present, it pushes the threshold for ignition to higher densities as discussed below.  


{\it Fusion ignition simulations:} We now perform thermal diffusion simulations of ignition.  Assuming constant pressure $P$, the conservation of energy can be written \cite{1992ApJ...396..649T},  
\begin{equation}
\frac{\partial E}{\partial t}+ P\frac{\partial}{\partial t}(\frac{1}{\rho_b})=\frac{1}{\rho_b}{\bf\nabla}\cdot \sigma{\bf \nabla} T + \dot S_{tot}\, ,
\label{Eq.E}
\end{equation}
where $E$ is the internal energy per baryon, $T$ the temperature, $\rho_b$ the baryon density, $\sigma$ the thermal conductivity and $\dot S_{tot}$ the total nuclear reaction heating rate per baryon.   We expect constant pressure to be a good approximation because the flame moves subsonically and sound waves can restore the background pressure.

Our goal is to demonstrate the physics in as clear a way as possible.  The simple equation of state we use is a largely degenerate very relativistic electron gas with internal energy per baryon,
\begin{equation}
E\approx Y_e\bigl(\frac{3}{4}\epsilon_F+\frac{\pi^2}{2}\frac{T^2}{\epsilon_F}\bigr)\, ,
\end{equation}
Fermi energy $\epsilon_F=(3\pi^2Y_e\rho_b)^{1/3}$, and electron fraction $Y_e$.  We use units $\hbar={\rm c}={\rm k}_b=1$.
The sum of the two terms on the left hand side of Eq. \ref{Eq.E} can be combined using the heat capacity at constant pressure $C_p=5\pi^2Y_eT/(4\epsilon_F)$, so that Eq. \ref{Eq.E} becomes,
\begin{equation}
\frac{\partial T}{\partial t}=\frac{1}{C_p\rho_b}{\bf\nabla}\cdot \sigma{\bf \nabla} T + \frac{\dot S_{tot}}{C_p}\, .
\label{Eq.T}
\end{equation}
We directly simulate Eq. \ref{Eq.T}, assuming spherical symmetry, with a simple first order implicit scheme \cite{koonin}.  The thermal conductivity $\sigma$ is from electron conduction where the mean free path is limited by electron ion scattering.  This scattering depends on the average charge $\langle Z\rangle$ of the ions which decreases as ions fission, see Fig. \ref{Fig2}. To evaluate $\sigma$ we use the simple formulas of \cite{1980SvA....24..303Y}.  The highly charged ions reduce $\sigma$ and somewhat slow thermal diffusion.  The initial conditions involve an actinide rich crystal for $0\le r\le r_{pit}$ and a 50/50\% (by mass) C/O liquid for $r_{pit} < r\le r_{grid}$ (except where O/Ne/Mg composition is noted).  Typically $r_{grid}= 2\times 10^{-3}$ to $4\times 10^{-3}$ cm.  The initial temperature $T_i$ is uniform across the grid and equal to the crystallization temperature of the actinide mixture $\approx3$  keV.  The boundary conditions are $\partial T(r,t)/\partial r|_{r=0}=0$ and $T(r_{grid},t)=T_i$.  Simulations typically use a time step of $10^{-15}$ s and a uniform grid spacing of $2\times10^{-6}$ cm.  Simulations with smaller time step and or grid spacing often yield very similar results.

The nuclear heating $\dot S_{tot}=\dot S_{fis}+\dot S_{fus}$ comes from both fission $\dot S_{fis}$ and fusion $\dot S_{fus}$.   For $r\le r_{pit}$, $\dot S_{fis}$ is taken from fission reaction simulations such as shown in Figs. \ref{Fig1},\ref{Fig2}.  Since the fission chain reaction does not depend strongly on temperature, nuclear network simulations are run first and then the results simply used in thermal diffusion simulations.  We note the total fission energy $S_{fis}$ for the simulation in Fig. \ref{Fig2} is $\int dt \dot S_{fis}(t)=0.679$ MeV/baryon.

To estimate $\dot S_{fus}$ \added{we calculate the rate of $C+C$ fusion using the REACLIB database \cite{Cyburt_2010} and include strong screening.   Using the rate from \cite{PhysRevC.74.035803} instead yields a slightly higher threshold density for carbon ignition.}  $C+C$ fusion produces a number of reaction products and these undergo secondary reactions. Careful reaction network simulations in \cite{Calder_2007} determined the total energy released during carbon burning to be $\Delta E=3.5\times 10^{17}$ ergs/g or 0.362 MeV/baryon at $8\times 10^8$ g/cm$^3$ (P200 network in Table 2 of \cite{Calder_2007} ).  For simplicity we calculate $\dot S_{fus}$ by assuming each $C+C$ fusion releases $Q_{eff}=(24/0.5)\Delta E=17.4$ MeV.  Here there are 24 nucleons per $C+C$ fusion and only 0.5 of the fuel is carbon.  This approximation can be checked by full reaction network simulations.  If $Q_{eff}$ is somewhat smaller, the threshold density for ignition may increase somewhat. 

\begin{figure*}[tbh]
\centering  
\includegraphics[width=\textwidth]{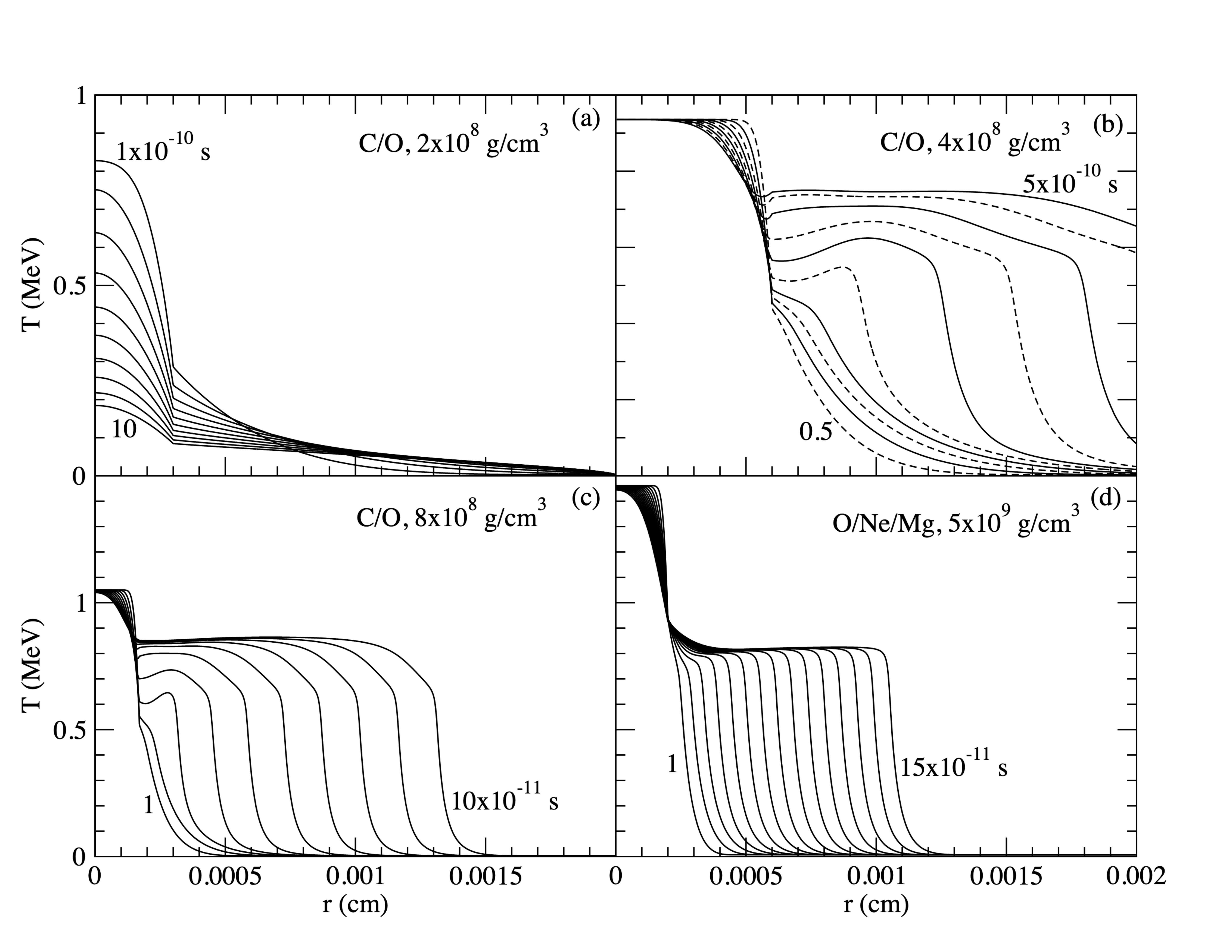}
\caption{\label{Fig3} Temperature $T$ versus radius $r$ of thermal diffusion simulations.  (a) Low density simulation that fails to ignite.  Contours are shown every $10^{-10}$ s top to bottom.  (b) Medium density simulation with a large $r_{pit}=6\times10^{-4}$ cm.  Contours are shown every $0.5\times10^{-10}$ s bottom to top.  (c) Simulation with $r_{pit}=1.7\times 10^{-4}$cm.   Contours are shown every $10^{-11}$ s left to right.  (d) O/Ne/Mg simulation at high density.  Contours are shown every $10^{-11}$ s left to right. \explain{Figure revised to now show results for REACLIB C+C and O+O rates.}}	
\end{figure*}

  
\section{Results}\label{Sec.Results}
Figure \ref{Fig3} shows temperature $T$ versus radius $r$ for four thermal diffusion simulations.  The simulation at a low density of $2\times 10^{8}$ g/cm$^3$ shown in Fig. \ref{Fig3} (a) fails to ignite.   Here $r_{pit}=3\times10^{-4}$ cm. During the fission chain reaction $T$ rises so rapidly that there is only minimal thermal diffusion.  However over longer times this heat simply diffuses away without initiating carbon burning.

At a density of $4\times 10^8$ g/cm$^3$, ignition is possible if $M_{pit}$ is large.  This is shown in Fig. \ref{Fig3} (b)  where a carbon flame is started that burns to the right (off the edge of the figure).   However $r_{pit}=6\times10^{-4}$ cm and $M_{pit}=0.36$ g.   This is larger than the 10-20 mg suggested from Table \ref{Table2}.   We conclude that ignition may be possible at this density, but only if $M_{pit}$ is large.  If $r_{pit}$ is much less than $6\times10^{-4}$ cm the simulation fails to ignite.

Figure \ref{Fig3} (c) shows carbon ignition for a simulation with $r_{pit}=1.7\times 10^{-4}$ cm at $\rho=8\times10^8$ g/cm$^3$.  At this density $M_{pit}=16$ mg is consistent with Table \ref{Table2} so ignition may be likely.  
If $r_{pit}$ is somewhat larger than $1.7\times 10^{-4}$ cm, ignition can take place at somewhat lower densities approximately $\ge 6\times 10^8$ g/cm$^3$.




The nuclear fission chain reaction in Figs. \ref{Fig1},\ref{Fig2} emits a total fission heating of 0.679 MeV per nucleon.  The fission energy released could be less if a smaller fraction of the actinides fission.  Alternatively, non-fissioning impurities such as Pb could be present that dilute the fission energy over more nucleons.  To explore this we multiply the fission heating rate in Fig. \ref{Fig2} by different time independent constants and find the total fission energy necessary for ignition at a given density, see Table \ref{Table3}.  The fission energy required decreases from 0.66 to 0.19 MeV/nucleon as the density increases from $6\times 10^8$ to $4\times 10^9$ g/cm$^3$.  For simplicity all of the simulations on which Table \ref{Table3} is based used $r_{pit}=3\times10^{-4}$ cm.

\begin{table}[tbh]
\caption{\label{Table3} Minimum  total fission heating $S_{fis}$ necessary for carbon ignition at a given density $\rho$. \explain{Table revised to now show results for REACLIB C+C rates.}}
\begin{tabular*}{0.272\textwidth}{c c } \hline
$\rho$ (g/cm$^3$)& $S_{fis}$ (MeV/nucleon)\\ \hline
$6\times10^8$& 0.66\\ 
$8\times10^8$&0.53\\
$1\times10^9$&0.46\\
$2\times10^9$&0.29\\
$3\times10^9$&0.22\\
$4\times10^9$&0.19\\
\hline \hline
\end{tabular*}
\end{table}

Oxygen ignition is difficult but appears possible at high densities.  Figure \ref{Fig3} (d), at a density of $5\times 10^{9}$ g/cm$^3$, shows oxygen ignition.   Again we use the REACLIB rates \cite{Cyburt_2010}. The initial composition is 60/30/10\% O/Ne/Mg by mass, $r_{pit}=2\times 10^{-4}$ cm, and $T_i=7$ keV.  We somewhat arbitrarily use an effective energy release of $Q_{eff}=16.4$ MeV per O+O fusion.  This is estimated from the Si rich final composition in Fig. 4b of \cite{1992ApJ...396..649T}.  Note that the system reaches a higher temperature $T\approx 1.5$ MeV after the fission chain reaction.  This is because, at very high densities, the system is more degenerate and the heat capacity is lower.  In all of the simulations  shown in Fig. \ref{Fig3}, the fission energy release is 0.679 MeV/nucleon.  It is possible that a very massive O/Ne star, near the Chandrasekhar mass, could experience a thermonuclear runaway via our mechanism.  In contrast an O/Ne WD might undergo electron capture induced collapse when it accretes matter from a companion.


\section{Discussion and Conclusions}\label{Sec.Conclusions}

{\it Electron capture and fission:} The threshold density for electron capture: $e+^{235}$U$\rightarrow ^{235}$Pa$+\nu_e$ is $9.2\times 10^7$ g/cm$^3$ (assuming $Y_e\approx 0.5$).  This may be followed by $e+^{235}$Pa$\rightarrow^{235}$Th$+\nu_e$ with a threshold of $2.0\times 10^8$ g/cm$^3$.  Thus the original $^{235}$U may be in the form of $^{235}$Th in the dense stellar interior.  $^{235}$Th with an even number of protons and an odd number of neutrons (like $^{235}$U) may be fissile and have a significant cross section for neutron induced fission.  In the laboratory $^{235}$Th beta decays so its fission cross section may not have been measured.  We note that the single neutron separation energy of $^{236}$Th is  5.9 MeV.  This is the energy available for $n+^{235}$Th fission and is significantly larger than the 4.8 MeV single n separation energy of $^{239}$U.   If $^{235}$Th does have a significant fission cross section, although somewhat smaller than for $^{235}$U, reaction network simulations such as in Fig. \ref{Fig1} still find comparable fission heating provided the initial $^{235}$Th enrichment compared to $^{238}$U+$^{235}$Th is somewhat higher than the 14\% assumed in Fig. \ref{Fig1}.

{\it Alpha decay lifetimes:} Uranium $\alpha$ decays with a 700 My half-life for $^{235}$U and 4.5 Gy for $^{238}$U.  However the $Q$ value (energy released) for $\alpha$ decay of $^{235}$Th is smaller than that for $^{235}$U so that the alpha decay systematics of \cite{VIOLA1966741} suggest $^{235}$Th will have a much longer half-life.  {\it In a dense star $^{235}$Th should be effectively stable}.  Over long time periods $^{238}$U will still decay.  As a result, the enrichment of $^{235}$Th compared to $^{238}$U will actually increase with time.  This could make a fission chain reaction more likely.

{\it Chandrasekhar limit:} The fission mechanism does not explicitly involve the Chandrasekhar mass limit.  Nevertheless, the high density required for ignition limits the mechanism to nearly Chandrasekhar mass WD and this might naturally produce transients of similar luminosities.  

{\it Ignition:} We have an explicit simulation of ignition.  We predict ignition at a single nearly central point in a very massive WD.  Nucleation of an actinide rich crystal is expected first in the highest density region and this should happen near, but perhaps not exactly at, the star's center.   Ignition takes place in a cold star.  Unlike in a conventional single degenerate model there is no period of carbon simmering before ignition.  Ignition produces a deflagration.  This might turn into a detonation later.  Hydrodynamic simulations of the SN or other astrophysical transient that might follow this cold ignition should be performed. \added{It is possible that these simulations, with ignition densities above $4\times 10^8$ g/cm$^3$, will reproduce reasonable typical SN Ia composition  \cite{2004NewAR..48..605T}.}

{\it Ultra-massive WD:} One way to form ultra-massive WDs with C/O cores is through mergers \cite{WDmerger}.  In our model the SN would not occur during or shortly after the merger.  Instead it would occur some time later when the massive star formed in the merger had cooled \cite{10.1093/mnras/stac348} so that actinide crystallization could start.  This is at about twice the temperature of C/O crystallization \cite{PhysRevLett.126.131101}.  This mechanism might be related to somewhat of a hybrid between single degenerate and double degenerate models.  Like the double degenerate model it would involve the merger of two WDs. Like the single degenerate model it would involve a deflagration ignition in a nearly Chandrasekhar mass WD.  An observable signature of our mechanism could be no detectable gravitational wave signal in a space based detector such as DECIGO \cite{https://doi.org/10.48550/arxiv.2006.13545} (because the merger happened in the past) along with the lack of an observable ex-companion star \cite{nocompanion}.  


In conclusion, we have performed thermal diffusion simulations of thermonuclear ignition following a natural nuclear fission chain reaction.  We find that carbon ignition is possible at high densities.  This could initiate a SN Ia or other astrophysical transient.


{\it Acknowledgements:} We thank Ed Brown, Ezra Booker, Alan Calder, Matt Caplan, Alex Deibel, 
Erika Holmbeck, Wendell Misch, Matthew Mumpower, Witek Nazarewicz, Rolfe Petschek, Catherine Pilachowski, Tomasz Plewa, 
and Rebecca Surman for helpful discussions.  This work was performed in part at the Aspen Center for Physics, which is supported by National Science Foundation grant PHY-1607611. This research was supported in part by the US Department of Energy Office of Science Office of Nuclear Physics grants DE-FG02-87ER40365 and DE-SC0018083 (NUCLEI SCIDAC).


\providecommand{\noopsort}[1]{}\providecommand{\singleletter}[1]{#1}%

\end{document}